\documentclass[fleqn,10pt]{wlscirep}
\usepackage{psfig}
\usepackage{amsmath}
\usepackage{subfigure}
\usepackage{hyperref}
\usepackage{authblk}
\usepackage{graphicx}

\usepackage{geometry}
\newcommand{\apj}{Astrophys. J.}

\newcommand{\aap}{Astron. \& Astrophys.}
\newcommand{\aj}{Astron. J.}
\newcommand{\mnras}{Mon. Not. R. astr. Soc.}
\newcommand{\pasp}{PASP}
\newcommand{\nat}{Nature}

\newcommand{\flux}{\,erg\,cm$^{-2}$\,s$^{-1}$}

\newcommand{\lum}{\,erg\,s$^{-1}$}

\newcommand{\cm}{\,cm$^{-2}$}
\newcommand{\nh}{$N_\mathrm{H}$}
\newcommand\arcdeg{\mbox{$^\circ$}}%
\newcommand\arcmin{\mbox{$^\prime$}}%
\newcommand\arcsec{\mbox{$^{\prime\prime}$}}%
\newcommand\farcs{\mbox{$.\!\!^{\prime\prime}$}}%
\newcommand{\alum}{\,\mathrm{erg\,s}^{-1}}

\title{A likely inverse-\textit{Compton} emission from the Type IIb SN 2013df}
\author[1,2,*]{K.~L. Li}
\author[1,*]{A.~K.~H. Kong}
\affil[1]{Institute of Astronomy and Department of Physics, National Tsing Hua University, Hsinchu 30013, Taiwan}
\affil[2]{Department of Physics and Astronomy, Michigan State University, East Lansing, MI 48824-2320, USA}
\affil[*]{\href{mailto:liliray@pa.msu.edu}{liliray@pa.msu.edu} (KLL) and \href{mailto:akong@phys.nthu.edu.tw}{akong@phys.nthu.edu.tw} (AKHK)}

\begin{abstract}
The inverse-\textit{Compton} X-ray emission model for supernovae has been well established to explain the X-ray properties of many supernovae for over 30 years. However, no observational case has yet been found to connect the X-rays with the optical lights as they should be. Here, we report the discovery of a hard X-ray source that is associated with a Type II-b supernova. 
Simultaneous emission enhancements have been found in both the X-ray and optical light curves twenty days after the supernova explosion. While the enhanced X-rays are likely dominated by inverse-\textit{Compton} scatterings of the supernova's lights from the Type II-b secondary peak, we propose a scenario of a high-speed supernova ejecta colliding with a low-density pre-supernova stellar wind that produces an optically thin and high-temperature electron gas for the Comptonization. 
The inferred stellar wind mass-loss rate is consistent with that of the supernova progenitor candidate as a yellow supergiant detected by the \textit{Hubble Space Telescope}, providing an independent proof for the progenitor. This is also new evidence of the inverse-\textit{Compton} emission during the early phase of a supernova. 
\end{abstract}
\begin{document}

\flushbottom
\maketitle
\thispagestyle{empty}

\section*{Introduction}

Supernova (SN) explosions produce strong X-ray emissions by the interactions between the fast outward-moving ejecta materials and the almost stationary circumstellar medium (CSM) deposited by the progenitors' stellar wind. The CSM would be heated up to $T\geq10^9$~K by the forward shock (often termed as circumstellar shock) in which free-free cooling and/or inverse-\textit{Compton} (IC) cooling are crucial to radiate the thermal energy out in X-rays within the first months \cite{1982A&A...111..140F,1996ApJ...461..993F}. For electron temperatures of the shocked CSM, $T_e\geq3\times10^9$, the effect of Comptonization on the X-ray emission becomes more important leading a domination over the free-free emission in 1--10~keV, even for a \textit{Compton}-thin electron gas (i.e., $\tau_e\ll1$) \cite{1982A&A...111..140F,1996ApJ...461..993F}. As the luminosity and the spectral slope of the Comptonized flux strongly depends on the optical depth that is totally determined by properties of the pre-supernova stellar wind, observing the post-supernova X-rays provides a unique way to trace the mass-loss history of the SN progenitor that gives constraint on the fundamental stellar properties. 

Supernova (SN) 2013df, residing in the constellation Coma Berenices, was discovered by the Italian Supernovae Search Project (ISSP) on 7.87 June 2013 UT with an apparent magnitude of 14th. 
Optical spectrum from the 10-meter Keck II telescope at Keck Observatory showed that it is a Type IIb SN with weak hydrogen emission features observed in the early phase \cite{2013CBET.3557....1C}. 
The host galaxy of SN 2013df, NGC~4414, was observed on 1999 April 29 by the \textit{Hubble} Heritage Project (GO/DD 8400; PI: Keith Noll) with the Wide Field and Planetary Camera 2 (WFPC2) of the \textit{Hubble Space Telescope} (HST) of which a three-band (F439W, F555W, and F814W) image mosaic was obtained in total fourteen WFPC2 pointings with a total exposure time of 5400 sec. 
By matching the WFPC2 images with a post-explosion 20 sec exposure taken by HST/WFC3 (F555W) on 2013 July 15, a yellow progenitor candidate within an astrometric accuracy of 5.5 mas was discovered \cite{2014AJ....147...37V}. 
Although the source was undetected in WFPC2/F439W band with a $3\sigma$ lower limit of 25.65~mag, they concluded that the candidate progenitor is likely a yellow supergiant with an effective temperature of $T_\mathrm{eff}=4250\pm100$ K, a bolometric luminosity of $L_\mathrm{bol}=10^{4.94\pm0.06}\,L_\odot$, and an effective radius of $R_\mathrm{eff}=545\pm65\,R_\odot$ by the absolute magnitudes of $V=-6.89$~mag and $I=-8.17$~mag. The authors also compared the maximum luminosity at the Type IIb secondary peak to those of the SNe IIb 1993J, 2008ax, and 2011dh and estimated less than $0.06M_\odot$ of $^{56}$Ni synthesized in the SN 2013df explosion. 

In addition to the discovery of the yellow supergiant progenitor, another interesting scientific finding for SN 2013df is the early X-ray detection by the \textit{Swift} X-ray Telescope (XRT) \cite{2013ATel.5150....1L}. In spite of the limited resolution of \textit{Swift}/XRT and the contamination from the diffuse emission of the host galaxy NGC~4414, the source is spectrally hard making it outstanding from the soft X-ray diffuse gas and showing the detection significance. 
Besides, early radio emissions were also detected by \textit{electronic Multi-Element Radio Linked Interferometer Network} (eMERLIN; at 5~GHz \cite{2015ATel.8452....1P}) and the \textit{Very Large Array} (VLA; at various frequencies from 1.5 to 43.7~GHz \cite{2015arXiv150407988K}). 
While it has been suggested that the radio and X-ray emissions primarily originated from synchrotron and bremsstrahlung, respectively, we here present the X-ray spectrum and the multi-wavelengths light curve of SN 2013df and argue the origins of the X-rays to be dominated by the \textit{Compton} scattering at the peak and the free-free emission in the rest of the time. In this work, the distance to NGC~4414 is assumed to be 16.6~Mpc \cite{2001ApJ...553...47F}. 

\section*{Results}

\subsection*{X-ray/Optical Light curves}

Double-peaked optical light curves are the most distinguished observational feature of Type IIb SNe. Shock wave heating of the progenitor's hydrogen envelope is the main cause of the first optical peak. While the amount of the hydrogen envelope (i.e., a few of 0.1 $M_\odot$ \cite{2011A&A...528A.131C}) is insufficient to support a sustainable thermal radiation, the decay of $^{56}$Ni synthesized in the explosion dominates the optical spectrum as the shock emission goes down, and produces the secondary optical peak \cite{1994ApJ...429..300W}. In the case of SN~2013df, the secondary optical peak showed up at about day 20 since the SN discovery \cite{2014MNRAS.445.1647M}. Surprisingly, a simultaneous X-ray brightening ($\sim4\sigma$ significance; see Methods for details) is also detected by \textit{Swift}/XRT (Figure \ref{fig:13df_lc}) with a photon index of $\Gamma=1.7^{+1.2}_{-1.0}$ and an X-ray luminosity of $L_\mathrm{1-7keV}\approx3.2\times10^{39}$\lum\ (1--7~keV; see Methods for details). From the \textit{Swift}/UVOT $v$-band and X-ray light curves, the X-ray brightening reached the maximum as the secondary optical emission peaked. In addition to the synchronicity of the emission peaks, the rising and decaying rate of the X-ray flux is about the same as those of the optical lights. It is presumably more than a coincidence and the X-ray emission should be closely related with the SN photospheric radiation. This is reminiscent of Inverse-\textit{Compton} scatterings of the SN lights by the shocked hot electron gas surrounding the SN as the source of the excess X-rays. 

\subsection*{IC scattering as the origin of the X-ray peak}

As mentioned in the introduction, IC is an important cooling process in the early phase of a SN, even if the shocked CSM electron gas (i.e., $T_e\geq10^9$~K) is \textit{Compton}-thin (i.e., $\tau_e\sim0.01$; see Methods for a detailed explanation). 
In fact, the IC X-ray spectrum can be described by a single power-law \cite{1982A&A...111..140F} and the spectral properties are closely related to the stellar wind (or the CSM) density of the progenitor, which can be represented by the ratio of the mass-loss rate (i.e., $\dot{M}_{-5}$ in units of $10^{-5}\,M_\odot\,$yr$^{-1}$) to the velocity (i.e., $w_{10}$ in units of 10~km~$s^{-1}$) of the pre-explosion wind. With the X-ray luminosity and the photon index obtained by \textit{Swift}/XRT, we apply the inverse-\textit{Compton} scattering model for SNe \cite{1982A&A...111..140F,1996ApJ...461..993F,1984A&A...133..264F} to explain the peak X-ray emissions observed in SN~2013df and constrain the wind parameters of the progenitor in the following (see Methods for a detailed description). 

As the photons of the secondary SN peak are the seeds of the IC X-rays, the temperature and bolometric luminosity of the SN are both important to affect the IC emission. Adopting the blackbody temperature of the secondary peak of $T_c\approx6900$~K \cite{2014MNRAS.445.1647M} and the absolute $v$-band magnitude of $M_v=-16.80$~mag ($D=16.6$~Mpc; see the Methods section), we estimated the bolometric luminosity, hence the radius of the SN photosphere, which is about $R_p \approx1.4\times10^{15}$ cm $\approx 2.1\times10^4\,R_\odot$. With an assumption of a typical SN shock velocity $V_s=10^4$~km~s$^{-1}$ (i.e., $T_e=1.36\times10^9$~K; see Methods) and the equations from eq.~(\ref{eq:opt_dep_ic}) to eq.~(\ref{eq:lum_ic}) for the optical depth, the photon index, and the IC luminosity, we computed the photon indices of the IC emission and the corresponding X-ray luminosities (i.e., 1--7~keV) with different wind parameters $\dot{M}_{-5}/w_{10}$ (a SN age of 20 days is assumed), which are shown with the best-fit XRT spectral values in Figure \ref{fig:13df_model}. Comparing with the \textit{Swift}/XRT data, it is clear that $\dot{M}_{-5}/w_{10}\approx4$ matches well with the best-fit results. The corresponding optical depth is $\tau_e\approx0.04$, which fulfils the minimum criterion of $\tau_e>0.01$. In addition, 
we also used the UVOT color temperature of the secondary peak (i.e., $T_c\approx5500$~K; see the Methods section) for the same analyses and the result is similar to the previous one with a slightly larger $\dot{M}_{-5}/w_{10}\approx5$ ($\tau_e\approx0.05$). Therefore, we conclude that the wind parameter of the progenitor is approximately $\dot{M}_{-5}/w_{10}\approx4-5$. 


\subsection*{A free-free emission as the ``quiescent'' component}

We attribute the so-called ``quiescent'' off-peak X-ray emission component to be a free-free emission of the shocked hot gas as indicated by the extremely flat X-ray spectrum observed (i.e., $\Gamma\approx1.0$, equivalent to a zero spectral index; See Methods for details). According to Fransson et al. \cite{1996ApJ...461..993F}, the free-free luminosity density is also a strong function of $\dot{M}_{-5}/w_{10}$ and a simplified version for $V=10^4$~km~s$^{-1}$ around 1~keV can be written as

\begin{eqnarray}
L_\mathrm{ff} (1\,\mathrm{keV}) = 4.9\times 10^{38}\,\zeta\,\bigg(\frac{\dot{M}_{-5}}{w_{10}}\bigg)^2\,\bigg(\frac{t}{1\,\mathrm{day}}\bigg)^{-1}\,\alum\,\mathrm{keV}^{-1}. 
\label{eq:Lff_ic}
\end{eqnarray}

By adopting $\zeta=0.86$ (see Methods) and $\dot{M}_{-5}/w_{10}=4-5$, the free-free X-ray luminosities (i.e., 1--7~keV and $\Delta L_\mathrm{ff}=6$~keV) in the time interval of $t=10$~d to 60~d are in the range of $6\times10^{39}$ to $0.7\times10^{39}$\lum, which are in a good agreement of the \textit{Swift}/XRT data. 

\section*{Discussion}

Chevalier \cite{1982ApJ...259..302C} suggested that such \textit{Compton} scattering radiations should dominate the UV band, equivalent to an energy range around 10--100~eV, while emissions in the X-ray band (i.e. 0.1--10~keV) are mainly caused by free-free scatterings. However, the Comptonization flux could sometimes be compatible to the free-free radiations in soft X-ray band; for instance, the luminosity ratio of the \textit{Compton}-to-free-free processes of SN~1993J (i.e., $L_\mathrm{Comp}/L_\mathrm{ff}$) in ROSAT range (i.e., 0.1-2.4~keV) for $\tau_e\approx0.05$ and $T_e\approx10^9$~K was computed to be 1.36 at day 7 \cite{1996ApJ...461..993F} and the ratio could increase to $L_\mathrm{Comp}/L_\mathrm{ff}=341$ with an electron temperature of $T_e=5\times10^9$~K. Although the ROSAT range is slightly different from the \textit{Swift}/XRT range used here, the electron scattering optical depth as well as the shock temperature are indeed very similar to what inferred from the \textit{Swift}/XRT peak emission spectrum. This ensures the possibility of the IC scattering scenario in such an optically-thin hot gas. 
Moreover, the inferred $\dot{M}_{-5}/w_{10}\approx4-5$ of SN~2013df is indeed very close to the one of SN 1993J inferred with the early X-ray data (i.e., $\dot{M}_{-5}/w_{10}\approx4$ \cite{1996ApJ...461..993F}). While the SN 1993J progenitor has been firmly characterized as an early K-type supergiant star \cite{2002PASP..114.1322V}, the similarity between the pre-explosion stellar winds of the two SNe suggests that the SN 2013df progenitor was also a yellow supergiant, supporting the cool supergiant detected by the  HST \cite{2014AJ....147...37V} as the true progenitor. 

\subsection*{Other Possible Mechanisms}
Synchrotron radiation is another possible mechanism to produce non-thermal X-rays in SNe. However, the expected X-ray emission inferred from the observed synchrotron radio spectrum is at least an order of magnitude fainter than what we have seen in SN~2013df \cite{2015arXiv150407988K}. 
We also considered an inverse \textit{Compton} emission of the synchrotron radio photons as the origin of the observed X-rays, but in this case the X-ray-to-radio ratio would be too large (i.e.,  the radio luminosity at 5~GHz is $L_R=2.6\times10^{36}$\lum\ \cite{2015arXiv150407988K} and $L_X/L_R>2000$) comparing with other well-studied SNe (e.g., $L_\mathrm{IC}/L_\mathrm{synch}<40$ for the SN~1998bw \cite{1998Natur.395..663K}). Therefore, both the above two mechanisms are not applicable in the case of SN~2013df. 

Alternatively, it has been suggested that the X-rays of SN~2013df are dominated by the bremsstrahlung thermal emissions from either the forward shock ($T_\mathrm{fs}\geq10^9$~K) or the reverse shock ($T_\mathrm{rs}\geq10^7$~K) \cite{2015arXiv150407988K}. 
Although we agree that the free-free emissions significantly contribute the observed X-rays in the early phase (see the previous section), bremsstrahlung alone is insufficient to explain the correlation between the optical/X-ray light curves. 
While the X-ray photon index of SN~2013df is harder than it should be in a Comptonization dominant case ($\Gamma_\mathrm{ph}\approx3$; see Figure \ref{fig:13df_model}), a hybrid of thermal bremsstrahlung and Inverse-\textit{Compton} is a good way to understand the multi-wavelength data as we suggested. 

\subsection*{Importance of the Inverse-\textit{Compton} emission for SN studies}
Inverse-\textit{Compton} model has been well established to explain X-ray emissions of various SN events, for examples, the Type II-L SNe 1979C, 1980K, the famous Type II-b SN 1993J, and the Type II-P SN~2006bp \cite{1982ApJ...259..302C,1982A&A...111..140F,1996ApJ...461..993F,2007ApJ...664..435I}. 
In fact, the IC scenario is also promising to explain the mysterious X-ray emissions of the special class GRB-SN associations (i.e., hydrogen-stripped Type Ic and possibly Ib supernovae detected well after the associated GRBs) and SN shock breakout events \cite{2008Natur.453..469S,2007ApJ...667..351W,2012grbu.book..169H}. 
While an IC X-ray brightening has been theoretically shown possible during the optical peaks of the SN \cite{2004ApJ...605..823B,2006ApJ...651..381C}, it has been a missing piece before for a long time. 
Although the recent multi-wavelength study of the Type Ia SN 2011fe (a.k.a., PTF 11kly) in M101 placed a marginal constraint on such a correlation by demonstrating a correlated X-ray/optical light curve profiles \cite{2012ApJ...751..134M}, the X-rays of SN 2011fe are actually undetected (i.e., the X-ray profile was constructed based on three upper-limits) and the corresponding flux upper limits place a 3$\sigma$ constraint on the progenitor mass loss rate of $\dot{M}_{-5}/w_{10}<2\times10^{-5}$ (note: it could be an interstellar stellar medium instead), which just ambiguously restricts the X-ray/optical relation. The X-ray brightening of SN 2013df shows a clear correlation with the secondary optical peak, which is an important example for the SN ejecta and CSM interaction model. 

The IC X-ray component study could also benefit various researches on Type Ia. As stated earlier, Type Ia supernovae are also possible IC X-ray emission sources, although no significant X-ray detection has been made from any known Type Ia (except for the SN 2014J in which $^{56}$Co lines at energies of 847 and 1238~keV were detected by INTEGRAL \cite{2014Natur.512..406C}). 
If IC scattering is significant, a few percent (i.e., $\sim\tau_e/2$) of the SN optical radiations will be IC scattered and the optical measurement will deviate from the true intrinsic SN brightness. 
As Type Ia supernovae are well-known as standard candles to measure cosmological distances by tracing accurately the unique SN decline, the measured brightness deviation could be important to fine-tune the distance ladder. IC emissions are also crucial to distinguish the proposed origins of Type Ia SNe, which are single-degenerate (a main-sequence star or red giant companion) and double-degenerate scenarios (another white dwarf companion), by examining density of the surrounding medium. 

On a conservative basis, the connection between the X-ray brightening of SN 2013df and the predicted IC X-ray emission may not be exclusively proven here. Nevertheless, this work certainly provides a new angle to the community on how simultaneous optical/X-ray data can be used to search for the IC emission of SNe. 
In particular, rapid \textit{Swift} multi-wavelength observations of SNe have been shown to be useful to such a search. 
Hopefully, \textit{Swift} will bring us more solid evidence of the IC emissions of SNe in the future.  

\section*{Methods}

\subsection*{\textit{Swift} X-ray Telescope (XRT)}

A series of \textit{Swift} ToO observations for SN 2013df was triggered since 2013 June 13 (6 days after the optical discovery) with a total number of 26 observations until 2013 August 6. From the combined 49 ks X-ray Telescope (XRT) observation, a hard X-ray source was detected at $\alpha\mathrm{(J2000)}=12^\mathrm{h}26^\mathrm{m}29^\mathrm{s}.326$, $\delta\mathrm{(J2000)}=31\arcdeg 13\arcmin 37\farcs87$ with an uncertainty of $3.6\arcsec$ (radius, 90\% confidence), which is less than $0.2\arcsec$ offset from the optical position. No significant X-ray source was found at the same position in the previous XMM-\textit{Newton} observations taken in 2005 -- 2006 with a 3$\sigma$ upper limit of $2.96\times10^{-15}$\flux\ or $9.8\times10^{37}$\lum\ (0.5--4.5~keV) deduced by the XMM-\textit{Newton} \texttt{FLIX} server. 

We downloaded the \textit{Swift} observations from the HEASARC archive and extracted XRT spectra and light curves using the \texttt{HEAsoft} (version of 6.11) built-in tasks, \texttt{xrtgrblc} and \texttt{xrtgrblcspec}. The source photons were selected using a circular region with a radius of $15\arcsec$ while a background subtraction was done using a source-free annulus region around the host galaxy. Unless otherwise mentioned, the uncertainties are in 90\% confidence level. 

\begin{figure}[ht]
\centering
\includegraphics[width=144mm]{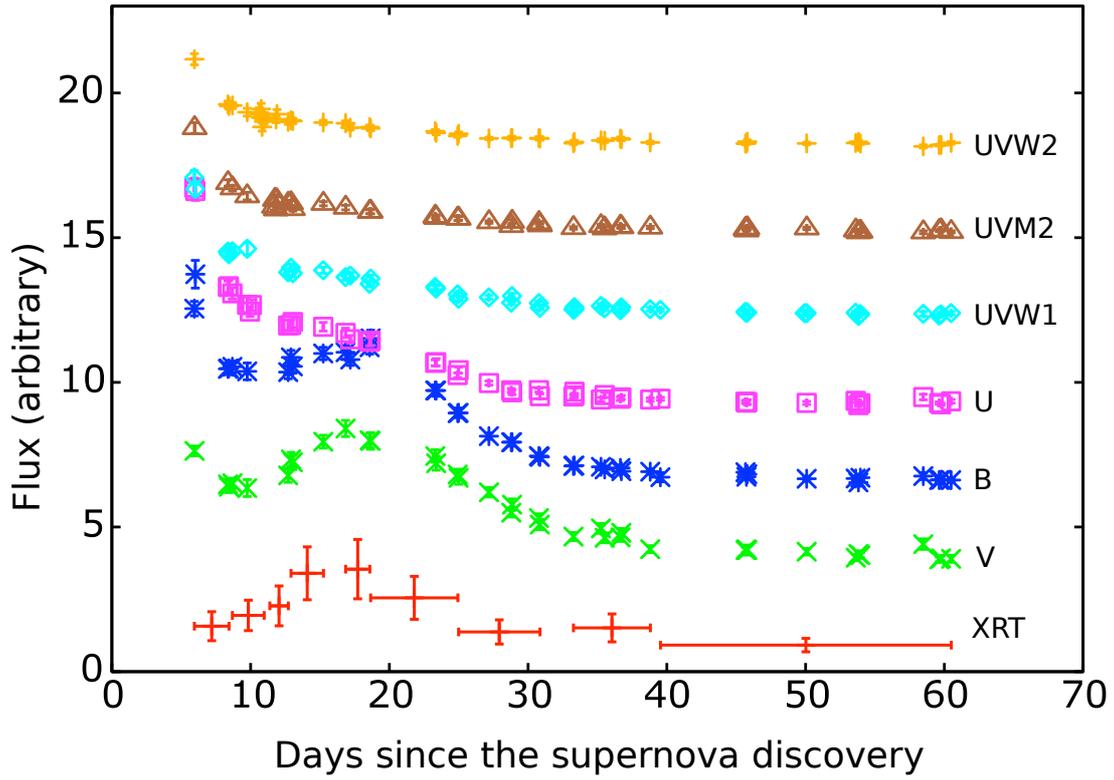}
\caption{Multi-wavelength light curves of SN~2013df with day zero fixed at the SN discovery. 
}
\label{fig:13df_lc}
\end{figure}

The \textit{Swift}/XRT light curve (Figure \ref{fig:13df_lc}; at least 10 counts per bin with \textit{Poisson} errors) shows that the X-ray emission lasted for about fifty-five days since the first \textit{Swift} observation, during which the X-rays gradually rises from $\sim5\times10^{39}$\lum\ at day 7 (with respect to the SN discovery here), reaches a peak luminosity of $\sim1\times10^{40}$\lum\ at about day 18, and drops steadily over the last 40 days of observations (Figure \ref{fig:13df_lc}). For the spectral model fitting, the XRT spectrum was binned with at least 15 counts per bin and weighted with the \textit{Gehrels} approximation \cite{1986ApJ...303..336G} (i.e., $\sigma\approx1+\sqrt{N+0.75}$ for a bin with $N$ counts) for a better uncertainty estimation in cases of $N<20$. 
We fitted the averaged ``hard'' XRT spectrum (i.e., 1--7~keV; to avoid possible contaminations from the soft X-ray diffuse emission of the host galaxy) with an absorbed power law model, of which the best-fit values are \nh\ $<2.6\times10^{22}$\cm\ and $\Gamma_\mathrm{ph}=0.91^{+1.4}_{-0.6}$ ($\chi_{\nu}^2=0.82$ with $dof=3$). In addition, we tried a thermal bremsstrahlung model to fit the data, but the best-fit plasma temperature hit the boundary (i.e., 200~keV) to fail the fit. Thus, we fixed the temperature to $kT=100$~keV (or $T_e\sim10^9$~K, a typical temperature of a free-free X-ray emission for supernovae \cite{1996ApJ...461..993F}), which cured the fit and gave \nh\ $<1.6\times10^{22}$\cm\ and $\chi_{\nu}^2=0.66$ ($dof=4$). Though a smaller $\chi_{\nu}^2$ value than that of the power-law model was obtained, we do not consider the improvement significant as the $\chi_{\nu}^2$ becomes 0.62 ($dof=4$) if we fixed the photon index to $\Gamma_\mathrm{ph}=1$. Instead, the simple power-law and the thermal bremsstrahlung models are both significant to describe the data and the former model fits the spectrum slightly better. 

\begin{figure}[ht]
\centering
\includegraphics[width=144mm]{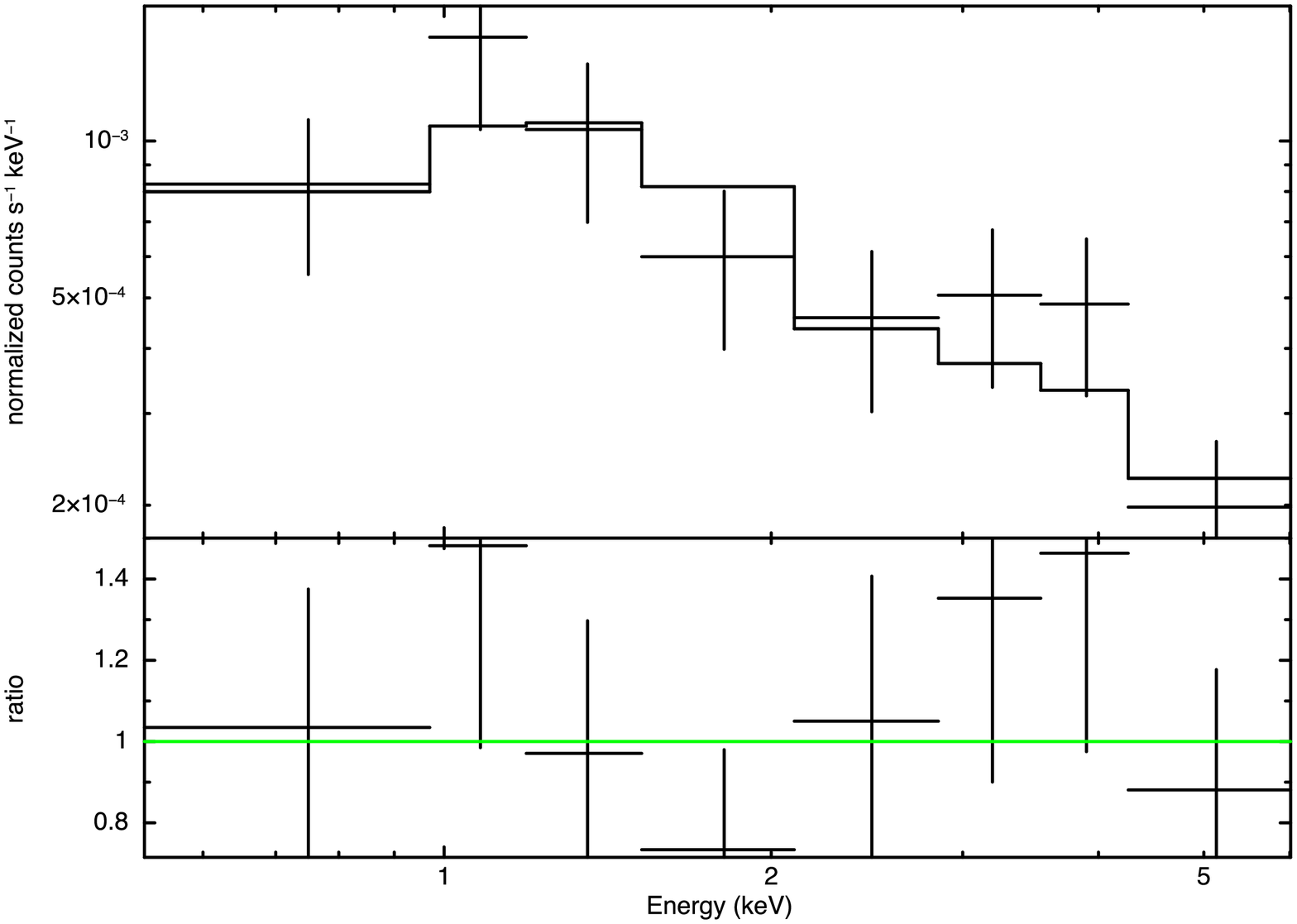}
\caption{\textit{Swift}/XRT spectrum of SN~2013df with the best-fit power-law of $\Gamma_\mathrm{ph}=0.96^{+0.50}_{-0.36}$ and \nh\ $<2.4\times10^{21}$\cm\ ($\chi_{\nu}^2=0.78$ with $dof=5$). }
\label{fig:pow_spec}
\end{figure}

As the photoelectric absorption is much more important below 1~keV, we extended the lower energy range of the XRT spectrum down to 0.3~keV to have a better constraint on the column density, hence to improve the fits. We also set the minimum allowed column density to the Galactic value \cite{2005A&A...440..775K} of \nh\ $=1.6\times10^{20}$\cm\ to prevent an non-logical null absorption. 
For an absorbed power law, the best-fit parameters are improved to $\Gamma_\mathrm{ph}=0.96^{+0.50}_{-0.36}$ and \nh\ $<2.4\times10^{21}$\cm\ ($\chi_{\nu}^2=0.78$ with $dof=5$; Figure \ref{fig:pow_spec}) yielding an unabsorbed luminosity of $L_\mathrm{0.3-7keV}\approx6.8\times10^{39}$\lum. 
For an absorbed thermal bremsstrahlung (temperature fixed to 100~keV), the best-fit \nh\ changes to \nh\ $<2.9\times10^{21}$\cm\ ($\chi_{\nu}^2=0.84$ with $dof=6$) and the luminosity is $L_\mathrm{0.3-7keV}\approx6.5\times10^{39}$\lum. 
Considering the poorly constrained \nh, we also computed the ``hard'' X-ray luminosity $L_\mathrm{1-7keV}\approx6.1\times10^{39}$\lum (power-law) and $L_\mathrm{1-7keV}\approx5.5\times10^{39}$\lum (bremsstrahlung), with which the influence of the greatly uncertain \nh\ on the investigation can be minimized. 

\subsection*{The X-ray brightening}

As mentioned earlier, an X-ray brightening is seen between day 12 and 25. We found that the brightening light curve can be described by a \textit{Gaussian} function on top of a flat light curve, $A\times\exp(-(t-t_p)^2/(2\sigma^2))+B$, where the the best-fit values are $A=(4.8\pm0.8)\times10^{-3}$~cts~s$^{-1}$, $t_p=17.1\pm0.7$~d (time at the peak), $\sigma=5.3\pm0.8$~d, and $B=(2.0\pm0.2)\times10^{-3}$~cts~s$^{-1}$ ($\chi^2=1.6; dof=5$). Alternatively, one may consider the apparent brightening as a statistical fluctuation. In this case, we have modelled the data with a flat light curve and the best-fit count rate is $(2.9\pm0.5)\times10^{-3}$~cts~s$^{-1}$ ($\chi^2=18.6; dof=8$). Clearly, there is a significant improvement (i.e., $\Delta\chi^2=17$) when considering the brightening a real detection (i.e., the \textit{Gaussian} model) rather than a fluctuation (i.e., the flat model). To find out whether this improvement is significant, we performed \textit{Monte Carlo} simulations by generating $10^6$ flat light curves and fitting them with both models. Among the $10^6$ trials, 141208 of them did not converge in the \textit{Gaussian} fit (i.e., a \textit{Gaussian} is totally unnecessary in these cases), 858765 of them have an improvement less than $\Delta\chi^2=17$, and only 27 trials were improved by $\Delta\chi^2\geq17$. This clearly shows that the X-ray brightening detection is not likely produced by chance (i.e., $p$-value $\approx0.003$\% or statistical significance $\sim4\sigma$). 

\begin{figure}[ht]
\includegraphics[width=144mm]{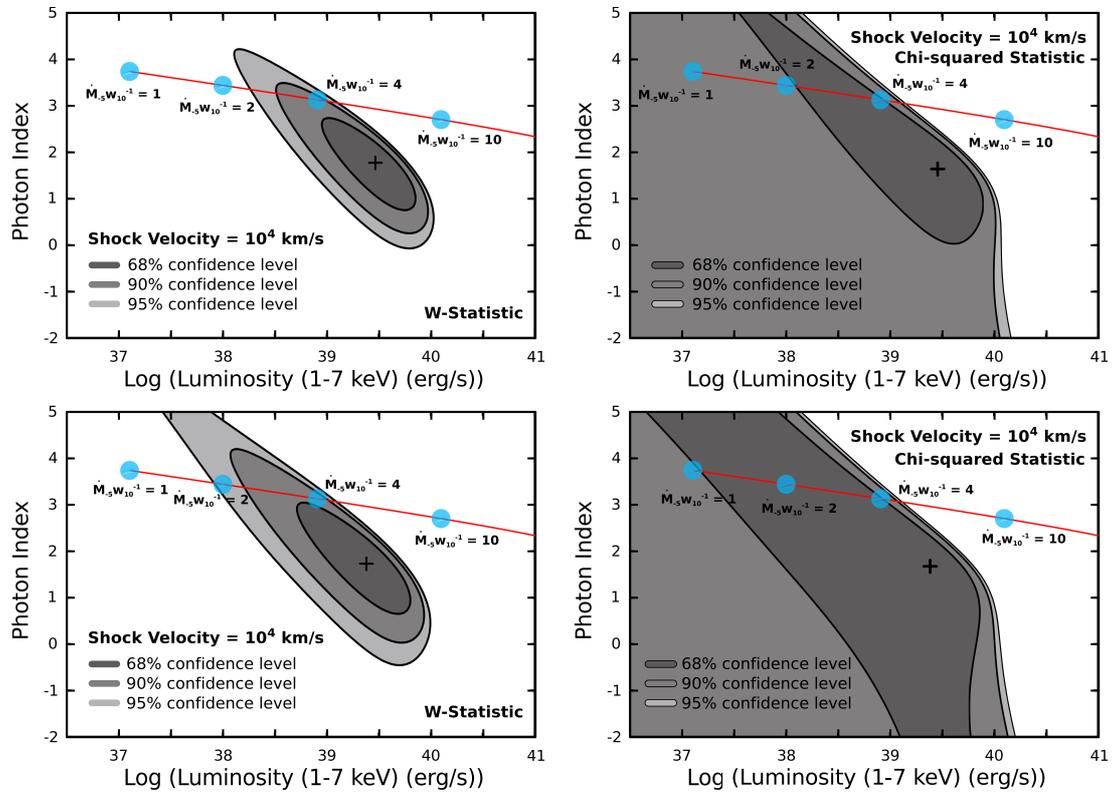}
\centering
\caption{Photon index versus logarithmic scale luminosity of the SN Comptonization model (red line; a SN photosphere of $T_c=6900$~K is assumed) plotted with contours of the best-fit at 68\%, 90\%, and 95\% confidence levels (grey shadows). The upper two plots are based on Approach \textit{A} while the lower two are based on Approach \textit{B} (see the text for details). }
\label{fig:13df_model}
\end{figure}

To investigate spectral features of the X-ray brightening, we used the observations of IDs from 32862008 to 32862015 to extracted a stacked X-ray peak spectrum. 
Given that the photon statistic is not good (i.e., spectral source counts: 28), we adopted a binning factor of at least 7 counts per bin to produce a 4-bin X-ray spectrum. 
To elaborate the small binning factor applied, it is a strategy that sacrifices the signal-to-noise (S/N) ratios of single bins to exchange for a larger number of degree-of-freedom. 
Otherwise, there will be only 1 or 2 bin(s) in the spectrum and leave no freedom for fitting. 
In fact, low-count binning is common in X-ray astronomy for faint sources (e.g., the \textit{Chandra} spectral analysis of A0620-00 \cite{2002ApJ...570..277K}). 
Additionally, we applied the \textit{W-statistic} (\textit{CASH-statistic} \cite{1979ApJ...228..939C} with \textit{Poisson} distributed background) to deal with the fitting process, instead of the standard \textit{Chi-squared statistic} (i.e., \textit{Gaussian} distribution is assumed), which does not work properly for low-count spectra (see the \href{https://heasarc.gsfc.nasa.gov/xanadu/xspec/manual/XSappendixStatistics.html}{XSPEC on-line manual} for details; the difference between \textit{W-statistic} and \textit{Chi-squared} performances is shown in Figure \ref{fig:13df_model}). 
This \textit{W-statistic} (or \textit{CASH-statistic}) has been widely used in spectral fitting with low-count X-ray spectra (e.g., the hyperluminous X-ray source candidate, CXO J122518.6+144545 \cite{2010MNRAS.407..645J,2015MNRAS.454L..26H}). It is therefore a valid statistical model for the cases of SN 2013df. 
Although our data have a low number of photon counts, the quality of the spectral fits can be reflected by the uncertainties of the estimated parameters given a proper statistic applied. We also note that spectral binning is not a necessary process for \textit{W-statistic} fitting. However, binned spectra provide $\chi^2$ values of the best fits (not the fitting statistic though) that are good indicators of the goodness-of-fit (i.e., one cannot tell the goodness-of-fit from the \textit{W-statistic}) to guide our data analysis. 

With the Galactic hydrogen column density of \nh\ $=1.6\times10^{20}$\cm\ assumed, the spectrum could be power-law distributed with a photon index of $\Gamma_\mathrm{ph}=1.4\pm0.6$ and the averaged luminosity of the X-ray peak is $L_\mathrm{0.3-7keV}\approx8.1\times10^{39}$\lum\ (or $L_\mathrm{1-7keV}\approx6.6\times10^{39}$\lum). Comparing with the ``quiescent'' spectrum extracted from the off-peak observations (i.e., Ob. ID: 32862001--00032862005, 32862016--00032862027; Table \ref{tab:swift_xrt}), of which the best-fit values are $\Gamma_\mathrm{ph}=1.0\pm0.4$ and $L_\mathrm{0.3-7keV}\approx4.9\times10^{39}$\lum\ (or $L_\mathrm{1-7keV}\approx4.4\times10^{39}$\lum), the peak X-ray emissions are spectrally softer, although insignificant. We also tried to subtract the peak spectrum by the ``quiescent'' one to probe for any additional emission component (i.e. IC emissions) leading to the X-ray enhancement as the observed peak. Two approaches have been used to eliminate the ``quiescent'' emission contribution, either by the observed spectrum itself (i.e., serving the ``quiescent'' spectrum as the ``background'' spectrum; namely approach \textit{A}) or the best-fit power law model (i.e., including the fixed best-fit ``quiescent'' spectral model of $\Gamma_\mathrm{ph, fix}=1.0$ and $\mathrm{norm_{fix}}=1.4\times10^{-5}$ photons keV$^{-1}$cm$^{-1}$s$^{-1}$ at 1 keV, to the model of interest; approach \textit{B}). 
For the approach \textit{A}, the subtracted spectrum can be described with a photon index of $\Gamma=1.7^{+1.2}_{-1.0}$ and an X-ray luminosity of $L_\mathrm{0.3-7keV}\approx4.3\times10^{39}$\lum\ (or $L_\mathrm{1-7keV}\approx3.2\times10^{39}$\lum). For the approach \textit{B} (i.e., two power-law components of which the ``quiescent'' one has been entirely frozen), the best-fit photon index and X-ray luminosity are $\Gamma=1.7^{+1.5}_{-1.2}$ and $L_\mathrm{0.3-7keV}\approx3.6\times10^{39}$\lum\ (or $L_\mathrm{1-7keV}\approx2.5\times10^{39}$\lum), respectively. While the fitting results of both methods are consistent with each other, the approach \textit{A} generally performs better than the approach \textit{B} in terms of the uncertainties of the best-fit parameters. 
Figure \ref{fig:13df_model} shows the contour maps of the best-fit photon indices and luminosities, which fully revealed their uncertainties. 
Combining with the IC emission spectral model for SNe, the maps are indeed very useful to constrain the wind parameter of the SN progenitor (i.e., $\dot{M}_{-5}/w_{10}\approx4-5$ for a shock velocity of $V_s=10^4$~km~s$^{-1}$). 

\subsection*{\textit{Swift} Ultraviolet and Optical Telescope (UVOT)}

\textit{Swift}/UVOT observed SN 2013df simultaneously with XRT using all six UVOT filters (central wavelengths are from 1928 to 5468$\mathrm{\AA}$). We used a \texttt{HEAsoft} task \texttt{uvotmaghist} to compute the magnitude/flux density of each UVOT image using an aperture radius of $3\arcsec$ (the optimum aperture size for UVOT images; see the \href{http://swift.gsfc.nasa.gov/analysis/threads/uvot_thread_aperture.html}{\textit{Swift}/UVOT on-line analysis thread} for details) with an annulus source-free background around the SN. 
A prominent double-peaked light curve feature of Type IIb is clearly shown in $v$ and $b$ bands, but marginally seen from $u$ to $um2$ bands and invisible in $uw2$ band (Figure \ref{fig:13df_lc}). 
The extinction corrected UVOT magnitudes of the first and second peaks are listed in Table \ref{tab:swift_uvot} with $A_v=0.30$~mag \cite{2014MNRAS.445.1647M}. Based on the relation between $B-V$ colors and color temperatures ($T_c$ in Kevin) for blackbody radiations (i.e., $B-V\approx(7090/T_c)-0.71$), inferred temperatures of the peaks are $T_c=8500\pm4000$~K and $T_c=5500\pm800$~K for the first and second peaks, respectively. These temperatures are roughly consistent with those obtained by optical spectroscopic fittings in Morales-Garoffolo et al. \cite{2014MNRAS.445.1647M} (i.e., $T_c=7700\pm500$~K at day 9 and $T_c=6900\pm500$~K at day 22). 

\subsection*{The IC emission spectral model for SNe}

According to Fransson et al. \cite{1996ApJ...461..993F}, the electron scattering optical depth behind the circumstellar shock is dependent of the shock velocity ($V_s$) and the stellar wind properties, which could be described as (note: it is a modified version for general uses as the original version is dedicated for the Type IIb SN 1993J, see Fransson et al. 1996 \cite{1996ApJ...461..993F} and references therein for details): 

\begin{eqnarray}
\tau_e = 2.314\times10^{-1}\,\zeta\,\bigg(\frac{\dot{M}_{-5}}{w_{10}}\bigg)\,\bigg(\frac{V_s}{10^4\,\mathrm{km\,s}^{-1}}\bigg)^{-1}\,\bigg(\frac{t}{1\,\mathrm{day}}\bigg)^{-1}. 
\label{eq:opt_dep_ic}
\end{eqnarray}

The $\dot{M}_{-5}/w_{10}$ is the key parameter for the stellar wind description, where $\dot{M}_{-5}$ is the mass-loss rate in units of $10^{-5}\,M_\odot\,$yr$^{-1}$ and $w_{10}$ is the wind velocity in units of 10~km~$s^{-1}$. There is another parameter $\zeta$, defined as $\zeta=(1+2\,n_\mathrm{He}/n_\mathrm{H})/(1+4\,n_\mathrm{He}/n_\mathrm{H})$ for describing the chemical composition of the wind and $n_\mathrm{He}/n_\mathrm{H}=0.1$ is assumed throughout this analysis (i.e., $\zeta=0.86$, which has been widely used to explain the X-ray emission of SN 1993J \cite{1996ApJ...461..993F}). In addition, the formula is constructed under an assumption of a steady stellar wind flow (i.e., $\rho_w\propto r^{-2}$), which is the simplest case of a circumstellar density profile. 
Fransson \cite{1982A&A...111..140F} has shown that there is a fraction of $\sim(\tau_e/2)^n$ of the supernova photospheric blackbody radiation up-scattered $n$ times by the energetic shock electrons to form high-energy X-ray emissions, even though the hot shock gas is optical thin to electron scattering. For optical depths smaller than $\tau_e\ll0.01$, each seed photon will at most be up-scattered only one time before it escapes the hot shock. In this case, the energies of the IC scattered photons will be doubled (i.e., $\Delta E/E\approx4kT_s/m_ec^2\approx1$ for $4kT_s\gg E$, with a typical shock temperature $T_s\approx10^9$~K) to possibly produce an excess UV tail at the shorter wavelength end of the blackbody spectrum, but certainly no X-rays in keV levels. For $\tau_e>0.01$, the number of electron scattering is large (i.e., $n>1$) and the \textit{Compton} cooling is therefore important. In this case, the Comptonized spectrum is a single power-law in X-rays (i.e., $N(E)\propto E^{-\gamma}$), with a photon index of 

\begin{eqnarray}
\gamma = (9/4-\ln((0.9228-\ln(\tau_e))\tau_e/2) /\alpha)^{1/2} - 1/2, 
\label{eq:index_ic}
\end{eqnarray}

where $\alpha=kT_s/m_ec^2$ (note: the $\gamma$ index is a spectral index (i.e., $F(E)\propto E^{-\gamma}$), instead of a photon index, in the original version \cite{1982A&A...111..140F}). The \textit{Compton} luminosity above 13.6~eV could be described by a semi-empirical expression as

\begin{eqnarray}
L_\mathrm{Comp} = 4.5\times10^{40}\,\bigg(\frac{R_p}{10^{15}\,cm}\bigg)\,\bigg(\frac{\tau_e}{0.01}\bigg)\,\bigg(\frac{T_\mathrm{eff}}{10^4}\bigg)^{\gamma+2}\,\bigg(\frac{T_s}{10^9~\mathrm{K}}\bigg)\,\alum, 
\label{eq:lum_ic}
\end{eqnarray}

where $R_p$ and $T_\mathrm{eff}$ are the radius and the temperature of the photosphere, respectively, and $T_s$ is the temperature of the shocked gas. According to Fransson et al. \cite{1984A&A...133..264F}, the shock temperature can be computed for a given shock velocity $V_s$ by 

\begin{eqnarray}
T_s = 1.36\times10^9\,\bigg(\frac{V_s}{10^4\,\mathrm{km\,s}^{-1}}\bigg)\,\mathrm{K}. 
\label{eq:Ts_ic}
\end{eqnarray}

The inferred shock temperature ($T_s$) is also the temperature of ions ($T_\mathrm{ion}$) in the shocked hot gas. As the equipartition time of the plasma is relatively short (i.e., about 1 day) in such a high density shocked region, the ion temperature is approximately the same as the electron temperature (i.e., $T_e\approx T_\mathrm{ion}$) \cite{1982A&A...111..140F}.

\section*{Acknowledgements}

This project was supported by the Ministry of Science and Technology
of Taiwan through grant 103-2628-M-007-003-MY3. Special thanks are given to Thomas P. H. Tam, Hsiang-Kuang Chang, Tomotsugu Goto, and Yi Chou for helpful discussions. 

\section*{Author contributions statement}

Data analysis: KLL; data interpretation: KLL and AKHK; manuscript: KLL and AKHK. 

\section*{Additional information}

The author(s) declare no competing financial interests. 

\begin{table}[ht]
\centering
\caption{\textit{Swift}/XRT Observation log of SN 2013df}
\begin{tabular}{lccc}
\hline
Date & Ob. ID & Exposure Time (s) & Note \\ 
\hline
5.94		&	32862001	&	2969.27 & Low	\\
8.36		&	32862002	&	1977.85 & Low	\\
9.95		&	32862003	&	139.86	 & Low	\\
8.68		&	32862005	&	1940.39 & Low	\\
10.62	&	32862006	&	3348.88 & Raising \\
10.62	&	32862007	&	489.50	 & Raising\\
11.69		&	32862008	&	497.00		& Peak	\\
11.69		&	32862009	&	2677.11		& Peak	\\
12.70	&	32862010	&	2310.00		& Peak	\\
15.23	&	32862011	&	1578.28		& Peak	\\
16.84	&	32862012	&	1523.35		& Peak	\\
18.57	&	32862013	&	1800.55		& Peak	\\
23.30	&	32862014	&	2170.14		& Peak	\\
24.92	&	32862015	&	1947.89		& Peak	\\
27.17	&	32862016	&	1573.29		& Low	\\
28.78	&	32862017	&	1997.83		& Low	\\
30.78	&	32862018	&	1840.50		& Low	\\
33.26	&	32862019	&	1353.55		& Low	\\
35.25	&	32862020	&	1221.18		& Low	\\
36.65	&	32862021	&	1558.31		& Low	\\
38.80	&	32862022	&	1156.25		& Low	\\
45.66	&	32862023	&	3638.54		& Low	\\
49.93	&	32862024	&	1086.33		& Low	\\
53.61	&	32862025	&	4060.58		& Low	\\
58.48	&	32862026	&	452.01		& Low	\\
59.62	&	32862027	&	3940.71		& Low	\\
\hline
\end{tabular}
\label{tab:swift_xrt}
\end{table}

\begin{table}[ht]
\centering
\small
\caption{\textit{Swift}/UVOT magnitudes of SN 2013df}
\begin{tabular}{lcccccc}
\hline
 & $v$ & $b$ & $u$ & $uvw1$ & $uvm2$ & $uvw2$ \\ 
\hline
1st Peak & 14.47 (0.04) & 14.59 (0.04) & 13.67 (0.05) & 14.09 (0.05) & 14.31 (0.05) & 14.77 (0.06)\\
2nd Peak & 14.30 (0.06) & 14.87 (0.06) & 14.67 (0.07) & 15.17 (0.06) & 15.58 (0.07) & ~$\cdots$~\\
\hline
\label{tab:swift_uvot}
\end{tabular}
\end{table}

\begin{thebibliography}{10}
\expandafter\ifx\csname url\endcsname\relax
  \def\url#1{\texttt{#1}}\fi
\expandafter\ifx\csname urlprefix\endcsname\relax\def\urlprefix{URL }\fi
\providecommand{\bibinfo}[2]{#2}

\bibitem{1982A&A...111..140F}
\bibinfo{author}{{Fransson}, C.}
\newblock \bibinfo{title}{{X-ray and UV-emission from supernova shock waves in
  stellar winds}}.
\newblock \emph{\bibinfo{journal}{\aap}} \textbf{\bibinfo{volume}{111}},
  \bibinfo{pages}{140--150} (\bibinfo{year}{1982}).

\bibitem{1996ApJ...461..993F}
\bibinfo{author}{{Fransson}, C.}, \bibinfo{author}{{Lundqvist}, P.} \&
  \bibinfo{author}{{Chevalier}, R.~A.}
\newblock \bibinfo{title}{{Circumstellar Interaction in SN 1993J}}.
\newblock \emph{\bibinfo{journal}{\apj}} \textbf{\bibinfo{volume}{461}},
  \bibinfo{pages}{993} (\bibinfo{year}{1996}).

\bibitem{2013CBET.3557....1C}
\bibinfo{author}{{Ciabattari}, F.} \emph{et~al.}
\newblock \bibinfo{title}{{Supernova 2013df in NGC 4414 = Psn
  J12262933+3113383}}.
\newblock \emph{\bibinfo{journal}{CBETs}}
  \textbf{\bibinfo{volume}{3557}}, \bibinfo{pages}{1} (\bibinfo{year}{2013}).

\bibitem{2014AJ....147...37V}
\bibinfo{author}{{Van Dyk}, S.~D.} \emph{et~al.}
\newblock \bibinfo{title}{{The Type IIb Supernova 2013df and its Cool
  Supergiant Progenitor}}.
\newblock \emph{\bibinfo{journal}{\aj}} \textbf{\bibinfo{volume}{147}},
  \bibinfo{pages}{37} (\bibinfo{year}{2014}).
\newblock 

\bibitem{2013ATel.5150....1L}
\bibinfo{author}{{Li}, K.~L.} \& \bibinfo{author}{{Kong}, A.~K.~H.}
\newblock \bibinfo{title}{{X-ray detection of SN 2013df}}.
\newblock \emph{\bibinfo{journal}{ATel}}
  \textbf{\bibinfo{volume}{5150}}, \bibinfo{pages}{1} (\bibinfo{year}{2013}).

\bibitem{2015ATel.8452....1P}
\bibinfo{author}{{Perez-Torres}, M.} \emph{et~al.}
\newblock \bibinfo{title}{{eMERLIN radio detection of SN2013df at 5.0 GHz}}.
\newblock \emph{\bibinfo{journal}{ATel}}
  \textbf{\bibinfo{volume}{8452}}, \bibinfo{pages}{1} (\bibinfo{year}{2015}).

\bibitem{2015arXiv150407988K}
\bibinfo{author}{{Kamble}, A.} \emph{et~al.}
\newblock \bibinfo{title}{{Radio and X-rays From SN 2013df Enlighten
  Progenitors of Type IIb Supernovae}}.
\newblock \emph{\bibinfo{journal}{ArXiv e-prints (submitted to \apj)}}
  \textbf{\bibinfo{volume}{arXiv:1504.07988}}, \bibinfo{pages}{1--15}
  (\bibinfo{year}{2015}).
\newblock 

\bibitem{2001ApJ...553...47F}
\bibinfo{author}{{Freedman}, W.~L.} \emph{et~al.}
\newblock \bibinfo{title}{{Final Results from the Hubble Space Telescope Key
  Project to Measure the Hubble Constant}}.
\newblock \emph{\bibinfo{journal}{\apj}} \textbf{\bibinfo{volume}{553}},
  \bibinfo{pages}{47--72} (\bibinfo{year}{2001}).

\bibitem{2011A&A...528A.131C}
\bibinfo{author}{{Claeys}, J.~S.~W.}, \bibinfo{author}{{de Mink}, S.~E.},
  \bibinfo{author}{{Pols}, O.~R.}, \bibinfo{author}{{Eldridge}, J.~J.} \&
  \bibinfo{author}{{Baes}, M.}
\newblock \bibinfo{title}{{Binary progenitor models of type IIb supernovae}}.
\newblock \emph{\bibinfo{journal}{\aap}} \textbf{\bibinfo{volume}{528}},
  \bibinfo{pages}{A131} (\bibinfo{year}{2011}).

\bibitem{1994ApJ...429..300W}
\bibinfo{author}{{Woosley}, S.~E.}, \bibinfo{author}{{Eastman}, R.~G.},
  \bibinfo{author}{{Weaver}, T.~A.} \& \bibinfo{author}{{Pinto}, P.~A.}
\newblock \bibinfo{title}{{SN 1993J: A Type IIb supernova}}.
\newblock \emph{\bibinfo{journal}{\apj}} \textbf{\bibinfo{volume}{429}},
  \bibinfo{pages}{300--318} (\bibinfo{year}{1994}).

\bibitem{2014MNRAS.445.1647M}
\bibinfo{author}{{Morales-Garoffolo}, A.} \emph{et~al.}
\newblock \bibinfo{title}{{SN 2013df, a double-peaked IIb supernova from a
  compact progenitor and an extended H envelope}}.
\newblock \emph{\bibinfo{journal}{\mnras}} \textbf{\bibinfo{volume}{445}},
  \bibinfo{pages}{1647--1662} (\bibinfo{year}{2014}).
\newblock 

\bibitem{1984A&A...133..264F}
\bibinfo{author}{{Fransson}, C.}
\newblock \bibinfo{title}{{Comptonization and UV emission lines from Type II
  supernovae}}.
\newblock \emph{\bibinfo{journal}{\aap}} \textbf{\bibinfo{volume}{133}},
  \bibinfo{pages}{264--284} (\bibinfo{year}{1984}).

\bibitem{1982ApJ...259..302C}
\bibinfo{author}{{Chevalier}, R.~A.}
\newblock \bibinfo{title}{{The radio and X-ray emission from type II
  supernovae}}.
\newblock \emph{\bibinfo{journal}{\apj}} \textbf{\bibinfo{volume}{259}},
  \bibinfo{pages}{302--310} (\bibinfo{year}{1982}).

\bibitem{2002PASP..114.1322V}
\bibinfo{author}{{Van Dyk}, S.~D.} \emph{et~al.}
\newblock \bibinfo{title}{{The Progenitor of Supernova 1993J Revisited}}.
\newblock \emph{\bibinfo{journal}{\pasp}} \textbf{\bibinfo{volume}{114}},
  \bibinfo{pages}{1322--1332} (\bibinfo{year}{2002}).

\bibitem{1998Natur.395..663K}
\bibinfo{author}{{Kulkarni}, S.~R.} \emph{et~al.}
\newblock \bibinfo{title}{{Radio emission from the unusual supernova 1998bw and
  its association with the {$\gamma$}-ray burst of 25 April 1998}}.
\newblock \emph{\bibinfo{journal}{\nat}} \textbf{\bibinfo{volume}{395}},
  \bibinfo{pages}{663--669} (\bibinfo{year}{1998}).

\bibitem{2007ApJ...664..435I}
\bibinfo{author}{{Immler}, S.} \emph{et~al.}
\newblock \bibinfo{title}{{X-Ray, UV, and Optical Observations of Supernova
  2006bp with Swift: Detection of Early X-Ray Emission}}.
\newblock \emph{\bibinfo{journal}{\apj}} \textbf{\bibinfo{volume}{664}},
  \bibinfo{pages}{435--442} (\bibinfo{year}{2007}).
\newblock 

\bibitem{2008Natur.453..469S}
\bibinfo{author}{{Soderberg}, A.~M.} \emph{et~al.}
\newblock \bibinfo{title}{{An extremely luminous X-ray outburst at the birth of
  a supernova}}.
\newblock \emph{\bibinfo{journal}{\nat}} \textbf{\bibinfo{volume}{453}},
  \bibinfo{pages}{469--474} (\bibinfo{year}{2008}).

\bibitem{2007ApJ...667..351W}
\bibinfo{author}{{Waxman}, E.}, \bibinfo{author}{{M{\'e}sz{\'a}ros}, P.} \&
  \bibinfo{author}{{Campana}, S.}
\newblock \bibinfo{title}{{GRB 060218: A Relativistic Supernova Shock
  Breakout}}.
\newblock \emph{\bibinfo{journal}{\apj}} \textbf{\bibinfo{volume}{667}},
  \bibinfo{pages}{351--357} (\bibinfo{year}{2007}).
\newblock 

\bibitem{2012grbu.book..169H}
\bibinfo{author}{{Hjorth}, J.} \& \bibinfo{author}{{Bloom}, J.~S.}
\newblock \emph{\bibinfo{title}{{The Gamma-Ray Burst - Supernova Connection}}},
  \bibinfo{pages}{169--190} (\bibinfo{publisher}{Cambridge University Press
  (Cambridge)}, \bibinfo{year}{2012}).

\bibitem{2004ApJ...605..823B}
\bibinfo{author}{{Bj{\"o}rnsson}, C.-I.} \& \bibinfo{author}{{Fransson}, C.}
\newblock \bibinfo{title}{{The X-Ray and Radio Emission from SN 2002ap: The
  Importance of Compton Scattering}}.
\newblock \emph{\bibinfo{journal}{\apj}} \textbf{\bibinfo{volume}{605}},
  \bibinfo{pages}{823--829} (\bibinfo{year}{2004}).
\newblock 

\bibitem{2006ApJ...651..381C}
\bibinfo{author}{{Chevalier}, R.~A.} \& \bibinfo{author}{{Fransson}, C.}
\newblock \bibinfo{title}{{Circumstellar Emission from Type Ib and Ic
  Supernovae}}.
\newblock \emph{\bibinfo{journal}{\apj}} \textbf{\bibinfo{volume}{651}},
  \bibinfo{pages}{381--391} (\bibinfo{year}{2006}).
\newblock 

\bibitem{2012ApJ...751..134M}
\bibinfo{author}{{Margutti}, R.} \emph{et~al.}
\newblock \bibinfo{title}{{Inverse Compton X-Ray Emission from Supernovae with
  Compact Progenitors: Application to SN2011fe}}.
\newblock \emph{\bibinfo{journal}{\apj}} \textbf{\bibinfo{volume}{751}},
  \bibinfo{pages}{134} (\bibinfo{year}{2012}).
\newblock 

\bibitem{2014Natur.512..406C}
\bibinfo{author}{{Churazov}, E.} \emph{et~al.}
\newblock \bibinfo{title}{{Cobalt-56 {$\gamma$}-ray emission lines from the
  type Ia supernova 2014J}}.
\newblock \emph{\bibinfo{journal}{\nat}} \textbf{\bibinfo{volume}{512}},
  \bibinfo{pages}{406--408} (\bibinfo{year}{2014}).
\newblock 

\bibitem{1986ApJ...303..336G}
\bibinfo{author}{{Gehrels}, N.}
\newblock \bibinfo{title}{{Confidence limits for small numbers of events in
  astrophysical data}}.
\newblock \emph{\bibinfo{journal}{\apj}} \textbf{\bibinfo{volume}{303}},
  \bibinfo{pages}{336--346} (\bibinfo{year}{1986}).

\bibitem{2005A&A...440..775K}
\bibinfo{author}{{Kalberla}, P.~M.~W.} \emph{et~al.}
\newblock \bibinfo{title}{{The Leiden/Argentine/Bonn (LAB) Survey of Galactic
  HI. Final data release of the combined LDS and IAR surveys with improved
  stray-radiation corrections}}.
\newblock \emph{\bibinfo{journal}{\aap}} \textbf{\bibinfo{volume}{440}},
  \bibinfo{pages}{775--782} (\bibinfo{year}{2005}).

\bibitem{2002ApJ...570..277K}
\bibinfo{author}{{Kong}, A.~K.~H.}, \bibinfo{author}{{McClintock}, J.~E.},
  \bibinfo{author}{{Garcia}, M.~R.}, \bibinfo{author}{{Murray}, S.~S.} \&
  \bibinfo{author}{{Barret}, D.}
\newblock \bibinfo{title}{{The X-Ray Spectra of Black Hole X-Ray Novae in
  Quiescence as Measured by Chandra}}.
\newblock \emph{\bibinfo{journal}{\apj}} \textbf{\bibinfo{volume}{570}},
  \bibinfo{pages}{277--286} (\bibinfo{year}{2002}).
\newblock 

\bibitem{1979ApJ...228..939C}
\bibinfo{author}{{Cash}, W.}
\newblock \bibinfo{title}{{Parameter estimation in astronomy through
  application of the likelihood ratio}}.
\newblock \emph{\bibinfo{journal}{\apj}} \textbf{\bibinfo{volume}{228}},
  \bibinfo{pages}{939--947} (\bibinfo{year}{1979}).

\bibitem{2010MNRAS.407..645J}
\bibinfo{author}{{Jonker}, P.~G.} \emph{et~al.}
\newblock \bibinfo{title}{{A bright off-nuclear X-ray source: a type IIn
  supernova, a bright ULX or a recoiling supermassive black hole in
  CXOJ122518.6+144545}}.
\newblock \emph{\bibinfo{journal}{\mnras}} \textbf{\bibinfo{volume}{407}},
  \bibinfo{pages}{645--650} (\bibinfo{year}{2010}).
\newblock 

\bibitem{2015MNRAS.454L..26H}
\bibinfo{author}{{Heida}, M.}, \bibinfo{author}{{Jonker}, P.~G.} \&
  \bibinfo{author}{{Torres}, M.~A.~P.}
\newblock \bibinfo{title}{{Discovery of a second outbursting hyperluminous
  X-ray source}}.
\newblock \emph{\bibinfo{journal}{\mnras}} \textbf{\bibinfo{volume}{454}},
  \bibinfo{pages}{L26--L30} (\bibinfo{year}{2015}).
\newblock 

\end{thebibliography}
\end{document}